\documentclass{ws-mpla}
\usepackage{graphicx}
\usepackage{xcolor}
\usepackage{subfigure}

\newcommand{\p}{\partial}
\newcommand{\bn}{\begin{equation}}
\newcommand{\en}{\end{equation}}
\def\({\left(}
\def\){\right  )}
\def\[{\left[}
\def\]{\right]}
 
\def\bny{\begin{eqnarray}}
\def\eny{\end{eqnarray}}
\begin{document}

\markboth{S Jamal}
{Quadratic Integrals}

\catchline{}{}{}{}{}

\title{Quadratic integrals of a multi-scalar  cosmological model
}

\author{\footnotesize SAMEERAH JAMAL}

\address{School of Mathematics,
University of the Witwatersrand, Johannesburg, South Africa.
Sameerah.Jamal@wits.ac.za}

\maketitle

\pub{Received (Day Month Year)}{Revised (Day Month Year)}

\begin{abstract}
In the context of FRW spacetime with zero spatial
curvature, we consider a multi-scalar tensor cosmology model under the pretext of obtaining quadratic conservation laws. We propose two new interaction
 potentials of the scalar field. Integral to this task, is the existence of dynamical Noether symmetries which are Lie-B\"{a}cklund transformations of the physical system. Finally, analytical solutions of the  field are found corresponding to each new model. In one of the models, we find that  the scale factor mimics $\Lambda$-cosmology in a special case.
\end{abstract}

\ccode{PACS Nos.: 98.80.-k, 95.35.+d, 95.36.+x.}
\section{Introduction}
Observational  cosmological data for the universe suggests  an accelerating
phase of expansion \cite{rie,per}. This phenomenon has posed many new problems. The source of late-time cosmic acceleration \cite{teg,kom,par} has been attributed to an unidentified  matter with a
negative  equation of state parameter, viz. dark energy. The cosmological constant, leading to $\Lambda$-cosmology,
is the simplest candidate for the dark energy. In the existing literature,  other cosmological models have been proposed to investigate the acceleration phase of the
universe. In most approaches,  a cosmic fluid is  introduced into Einstein’s General Relativity \cite{bar,rat,ben,pan}, or other studies
modify the Einstein-Hilbert Action \cite{fer,roy,sot}.
In this work we are interested in two interacting scalar fields in a spatially flat FRW spacetime.

One important spatially homogeneous scalar field model is that containing  exponential potential 
energy  \cite{hall,cole}. A number of 
 inflationary models that include scalar fields have been proposed \cite{olive}.  Such models have an important role in higher dimensional gravity
theories and   in studying the dynamics of the early universe.
In particular, power-law inflationary
models (important in solving the horizon and flatness problems) arise in models with a scalar field having an exponential potential \cite{wet}.
For single scalar fields, a very steep potential will
not cause late-time acceleration \cite{cope}, while for multi-scalar field models, the universe can evolve to
accelerate era, i.e. assisted inflation \cite{guo,cope2} and assisted quintessence \cite{kim}. We note that a large body of literature exists for the  analytic solutions of single scalar fields, for example,  \cite{mus,ellis,east,kim2}. {In relation to our study, we mention the Chiral cosmological model \cite{l1,l2} that have important applications in 
inflation \cite{l3} and descriptions for $\alpha$-attractors \cite{l4,l5}. }

Currently, there are two primary ways to identify the unknown potential function: a) selection  by reconstruction of  special cosmological behavior according to some well-known
model, or b) 
phenomenological observations from high energy physics. 
Nevertheless, the functional form of a potential cannot
be posed only at a phenomenological level but should also be justified 
rigorously at a mathematical  level. At this juncture, symmetry based methods are one such option. 
Indeed, the main motivation to apply Noether symmetry methods for cosmological models is that when a  potential function
is not  unique and is an unknown function, one is able to find the explicit
forms of the potential function which possesses a symmetry. Additionally, group invariant transformations  are vital tools in physical problems as they feature in multiple applications, namely, they may be used to reduce the order of a given dynamical system, construct conservation laws (if they satisfy Noether's theorem), assess the integrability of the
physical system and find exact solutions. Practically, there are two classes of Noether symmetries that are of importance. The first is a Noether point symmetry, where the vector field itself contains no derivative terms. The second type of symmetry is called a Noether
contact or dynamical symmetry. The latter are the
generators of  infinitesimal  Lie-B\"{a}cklund transformations - linear in first derivatives.  Noether point symmetries give rise to linear conserved charges \cite{hans} while in contrast, Noether contact symmetries 
admit quadratic conserved charges. Now, our interest lies in the contact symmetries and hence the quadratic conservation laws. As mentioned above, a significant by-product of this study will be the determination of the  interaction potential function.

When it comes to point symmetries, one finds their presence in cosmology in a number of different contexts. In particular, we refer the reader to \cite{kew,cap,cot,vak,p1,souza,san,b4,j10,sit,jgp,b5} for examples where such symmetries have played a role in model determination.
On the other hand, applications of contact symmetries in  cosmology can be found in, inter alia \cite{ros2,p2}. {In particular, Killing tensors have been applied to find quadratic conservation laws in $f(R)$ cosmology with an ideal gas \cite{l6}, Brans-Dicke cosmology \cite{l7}, $f(R)$ gravity \cite{l8}, a discussion in \cite{l9}, a generic model \cite{l10} and more recently, Brans-Dicke cosmology with a scalar field \cite{l11}.}

  Contact
transformations have considerable advantages over point symmetries in the realm of Newtonian physics and  General Relativity, i.e they are closely related to physically meaningful conservation laws such as the
Runge-Lenz Vector \cite{con}, the Lewis Invariant \cite{lew} and conserved quantities in the  MICZ Problem
\cite{zwan,mc}. Moreover, the Carter constant in  Kerr
spacetime \cite{cart} stems from dynamical symmetries.  Yet another property of the integrals of dynamical symmetries is that they possess
more degrees of freedom that allows one to consider numerous scenarios in a given dynamical problem.

In lieu of this discussion, we emphasize the fact that the determination of Noether contact symmetries  is simply a geometric problem. The solution of this problem is provided through differential geometry, which relates the Killing tensors of the metric of the space generated by the dynamic fields to the dynamical symmetries \cite{c3}.

In gravitational field theories, the key to this idea is that for some minisuperspace  defined by a cosmological model, the Killing tensors are constructed from the model itself. Hence, the  existence of Noether contact transformations from a cosmological Lagrangian facilitates a theoretical basis to select functional forms of the model - a selection that conforms to the  geometric character of gravity.
Another way to determine potential functions is to
include some ad hoc assumption, see 
 \cite{bert,frie,sah}. In most cases however, 
the complexity of the field equations is considerable and  substantial mechanisms for simplifying differential equations are required in the search for exact solutions of the cosmological
models.

Back to our study, for the underlying  problem, 
we define the scalar field interaction in the kinetic and potential part, and require that the field equations  are  invariant under  contact
transformations.  To begin, we shall explore the application of the Killing tensors of the minisuperspace which generate dynamical
symmetries (quadratic conservation laws) for the field equations. The requirement that the field equations admit dynamical symmetries linked to quadratic conservation laws  results in the discovery of
two potential functions. By implication, these fields are integrable.  We obtain the corresponding analytical solution of the field
equations by first transforming the fields using normal coordinates. Thereafter, for the second potential, we show that in a special case the scale factor resembles $\Lambda$-cosmology and follows an exponential law at late-time.

The structure of the paper follows. In section 2 we consider scalar fields interacting both 
kinematically and potentially in a spatially flat FRW spacetime. We present some of the elemental properties of the dynamical
system such as the Action, the Lagrangian and field equations. In Section 3, we very briefly
list all pertinent theory regarding Noether contact symmetries. Section 4 contains the Killing vectors and tensors of the metric space governing the field under study. In Section 4, we determine
the potentials of the fields that admit quadratic conservation laws. Section 5 discusses and solves the resulting field equations, and Section 6 concludes.

\section{The Cosmological Model}

We consider  a two-scalar field cosmological model in General Relativity with Action Integral \cite{tam,bel,crem}
\bn S=\int dx^4 \sqrt{-g}\left[ R-\frac12 g_{ij}H_{AB}\(\Phi^{C}\)  \Phi^{A,i}\Phi^{B,i}+V\(\Phi^C\) \right],\label{ac} \en
where $A,B,C=1, 2$, $H_{AB}$ describes the coupling between the two scalar fields and $\Phi^C=(w,z)$. 

Notably, the action (\ref{ac}) imbibes a multitude of
 alternative gravitational theories  under a conformal transformation
\cite{star}.
We adopt a  spatially flat FRW spacetime $diag\(-1,a^2(t)\Sigma^3\)$ for the interacting fields, where $a(t)$ is the scale factor. Here $\Sigma^3$ is the metric of three dimensional Euclidean space.
Hence, the point-like Lagrangian becomes 
\bn L\(a,\dot{a},\Phi^C,\dot{\Phi}^C\)=-3a\dot{a}^2+\frac12 H_{AB}\dot{\Phi}^A\dot{\Phi}^B-a^3V\({\Phi}^C\).\label{l0g} \en The Lagrangian is holonomic and it admits time translation as a trivial symmetry.
There are two considerations for the  2D symmetric metric $H_{AB}$: one may assume that $H_{AB}$ is flat \cite{jamil} and hence admits three Killing vectors which span the $E (2)$ group, or alternatively, 
that $H_{AB}$ is a space of constant non-vanishing curvature \cite{p3}, once again admitting three Killing vectors,  but span instead the $SO(3)$ group. In the work that follows, we opt to consider the latter case.

Suppose that $h_{AB}=\frac38 H_{AB}$, where $h_{AB}=diag\(1,e^{2z}\)$, and moreover let $a=\(\frac38\)^{\frac13}u^{\frac23}.$


Therefore the Lagrangian (\ref{l0g}) may be written as \cite{p3}
\bn L\(u,\dot{u},w,\dot{w},z,\dot{z}\)=-\frac12\,{\dot{u}}^{2}+\frac12\,{u_{{}}}^{2} \left( \dot{z}^{2}+{\dot{w}}^{
2}{\rm e}^{2\,z} \right) -u^2V(w,z).\label{l0}\en 
The field equations are the Hamiltonian \cite{p3}
\bn E=-\frac12\,{\dot{u}}^{2}+\frac12\,{u_{{}}}^{2} \left( \dot{z}^{2}+{\dot{w}}^{
2}{\rm e}^{2\,z} \right) +u^2V(w,z), \label{f11} \en and the Euler-Lagrange equations \cite{p3}
\bny
 \ddot{u}=-{{\rm e}^{2\,z_{{}}}}u_{{}}{\dot{w}}^{2}-u_{{}}\dot{z}^{2}+2\,u_{{}}V \left( z_{{}},w_{{}} \right) ,\nonumber\\
\ddot{w}=-{\frac 
{2\,{{\rm e}^{2\,z_{{}}}}u_{{}}\dot{w}\dot{z}+2\,{{\rm e}^{2\,z_{{}}}}
\dot{u}\dot{w}+ V_{,w}  u_{{}}}{{{\rm e}^{2\,z_{{}}}}u_{{}}}},\nonumber\\
\ddot{z}=-{\frac {-{{\rm e}^{2\,z_{{}}}}u_{{}}{\dot{w}}^{2}+V_{,z}
 u_{{}}+2\,\dot{u}\dot{z}}{u_{{}}}}. \label{f2} \eny

Eventually, we will show that adhering to the aforementioned discussion
results in two cases of the interaction potential. Given the importance of analytical solutions to understand the evolution of the universe, we will use 
the Noether vectors to lead us to the solution of the field equations.

\section{Dynamical Noether Symmetries}

There is a fundamental relationship between 
 Killing tensors and conservation laws. To view this relation, consider a particle moving in  $n$ dimensional Riemannian space with metric $g_{ij}\(x^k\)$
under the action of the
potential $V\(x^k\)$. Let $\mathcal{L} \(x^k,\dot{x}^k\)=\frac12 g_{ij}\dot{x}^i\dot{x}^j-V(x^k),$ where $\dot{x}^i=\frac{dx^i}{dt}$, be the
Lagrangian function that defines a given dynamical system. We define the quadratic function \bn I=X^i\(t,x^k,\dot{x}^k\)\frac{\p \mathcal{L}}{\p \dot{x}^i}-B\(t,x^i\),\label{i0}\en
as the conserved quantity for the dynamical system, that is $\frac{d}{dt} I=0$. $I$ is quadratic and  clearly time-independent.
The vector field $X^i\(t,x^k,\dot{x}^k\)\p_i$ that provides the quadratic conservation law (\ref{i0}) is called a dynamical or contact Noether symmetry for the Lagrangian $\mathcal{L}  \(x^k,\dot{x}^k\)$. If the dependence of the vector field $X$ excludes derivative terms such as $\dot{x}^k$, then $X$ is called a Noether point symmetry and in this case, the integral $I$ will be linear instead.
The functional $B\(t,x^i\)$ is a boundary term  introduced to allow for the infinitesimal changes in the value of the Action Integral. 

The dynamical symmetry is given by the expression \bn X^{[1]}\mathcal{L} =\dot{B},\label{con}\en where $X^{[1]}=X+\dot{X}^i\p_{i}$. Alternatively, if we suppose that \bn X=K^i_{~{j}}\(t,x^k\)\dot{x}^i\p_{i},\label{con4}\en then condition (\ref{con}) is equivalent to the following system \cite{c2,c3}  \bny K_{\(ij;k\)}=0,\label{con1}\\
K_{ij,t}=0,\quad B_{,t}=0,  \label{con2}\\
K^{ij}V_{,j}+B_{,i}=0,\label{con3}\eny
where `;' denotes the covariant derivative with respect to the connection coefficients of the metric $g_{ij}$. From the conditions (\ref{con2}), we have that $K^i_{~{j}}\(x^k\)$ and $B\(x^k\)$,  condition (\ref{con1}) implies that $K^i_{~{j}}\(x^k\)$ is a second-rank Killing tensor of the metric $g_{ij}$, and lastly, condition (\ref{con3}) can be seen as a constraint condition for the potential $V(x^k)$ in relation to the Killing tensor $K^i_{~{j}}$ and boundary term $B.$

The use of dynamical Noether symmetries provides first integrals which can
be used to reduce the order of the dynamical system and possibly lead to analytic solutions of the field equations. In the present paper, we are interested in contact transformations for which the generator $X$ comprises 
of a second-rank tensor that  is symmetric on the indices, that is, $K_{[ij]} = 0.$

\section{Killing Tensors and Killing Vectors}
The metric \bn ds^2=-du^2+u^2\(dz^2+e^{2z}dw^2\), \label{m1}\en
is obtained from the Lagrangian (\ref{l0}), and 
it is easy to show that the metric  is flat. Furthermore, (\ref{m1}) admits a seven dimensional homothetic Lie algebra consisting of
the following vector fields:

- Three gradient Killing vectors (translations) 

$$T^1=\({{\rm e}^{z}}{w}^{2}+{{\rm e}^{-z}}\)\p_u-2\,{\frac {w{{\rm e}^{-z}}}{u}}\p_w+{\frac {-{{\rm e}^{z}}{w}^{2}+{{\rm e}^{-z}}}{u}}\p_z,$$
$$T^2=-{{\rm e}^{z}}w\p_u+{\frac {{{\rm e}^{-z}}}{u}}\p_w+{\frac {{{\rm e}^{z}}w}{u}}\p_z,$$
$$T^3=-{{\rm e}^{z}}\p_u+{\frac {{{\rm e}^{z}}}{u}}\p_z.$$
-Three non-gradient Killing vectors (the rotations) which span the $SO (3)$ algebra

$$R^1=\p_w,~R^2=\p_z+w\p_w,~R^3=w\p_z-\frac12\(w^2-e^{-2z}\)\p_w,$$ and the homothetic vector $u\p_u.$

To solve the determining conditions (\ref{con1})-(\ref{con3}), it is necessary to obtain the second-rank symmetric Killing tensors of the metric (\ref{m1}).   Once this is established, we will define the potential functions $V\(w,z\)$ that
admit quadratic conservation laws. 

For the space of  dependent variables $\{u,w,z\}$, in order to obtain the dynamical symmetry (\ref{con4}), 
 we find that the general form of the symmetric second-rank Killing tensor is:
$$K_{ij}={\left(
\begin{array}{ccc}
 K^u_{~{u}}\(u,w,z\) & K^u_{~{w}}\(u,w,z\)  &K^u_{~{z}}\(u,w,z\)  \\
 K^u_{~{w}}\(u,w,z\)  & K^w_{~{w}}\(u,w,z\)  & K^w_{~{z}}\(u,w,z\)  \\
 K^u_{~{z}}\(u,w,z\)  &K^w_{~{z}}\(u,w,z\)  & K^z_{~{z}}\(u,w,z\) 
\end{array}
\right)},$$

where (all $c_i$'s are constants) we define
\bny K^u_{~{u}}\(u,w,z\)={{\rm e}^{2\,z}}{c_9}\,{w}^{4}-{{\rm e}^{2\,z}}{c_{10}}\,{w}^{
3}+\frac12\,{{\rm e}^{2\,z}}{c_{12}}\,{w}^{2}+{{\rm e}^{2\,z}}{c_{13}}\,w+2\,{c_9}\,{w}^{2}\nonumber\\
+{c_9}\,{{\rm e}^{-2\,z}}+{
{\rm e}^{2\,z}}{c_{14}}-{c_{10}}\,w-{c_{11}},\label{k1}\eny

\bny K^u_{~{w}}\(u,w,z\)=-\frac18\,u \Big( -2\,u{{\rm e}^{3\,z}}{w}^{4}{c_4}-2\,u{{\rm e}^{3
\,z}}{w}^{3}{c_5}+u{{\rm e}^{3\,z}}{w}^{3}{c_6}-16\,{{\rm e}
^{2\,z}}{c_9}\,{w}^{3}\nonumber\\
+4\,u{{\rm e}^{3\,z}}{c_{17}}\,{w}^{2}-6
\,u{{\rm e}^{z}}w{c_5}
-u{{\rm e}^{z}}w{c_6}+12\,{{\rm e}^{2
\,z}}{c_{10}}\,{w}^{2}-4\,u{{\rm e}^{3\,z}}{c_{18}}\,w+2\,u{c_4}\,{{\rm e}^{-z}}\nonumber\\+4\,u{{\rm e}^{z}}{c_{17}}-4\,{{\rm e}^{z}}{
c_7}\,u-4\,{{\rm e}^{2\,z}}{c_{12}}\,w
-4\,u{{\rm e}^{3\,z}}{
c_{19}}\nonumber\\-4\,{{\rm e}^{2\,z}}{c_{13}}-16\,{c_9}\,w+4\,{c_{10}} \Big),\nonumber\\\label{k2}\eny

\bny K^u_{~{z}}\(u,w,z\) =-\frac14\, \Big( 2\,u{{\rm e}^{z}}{c_4}\,{w}^{3}-4\,{{\rm e}^{2\,z}}
{c_9}\,{w}^{4}-u{{\rm e}^{z}}{w}^{2}{c_6}+4\,{{\rm e}^{2\,z}
}{c_{10}}\,{w}^{3}\nonumber\\+2\,{{\rm e}^{-z}}{c_4}\,uw-2\,u{{\rm e}^{z}
}{c_7}\,w
-2\,{{\rm e}^{2\,z}}{c_{12}}\,{w}^{2}+2\,{{\rm e}^{-z
}}{c_5}\,u-2\,u{{\rm e}^{z}}{c_8}\nonumber\\
-4\,{{\rm e}^{2\,z}}{c_{13}}\,w+4\,{c_9}\,{{\rm e}^{-2\,z}}-4\,{{\rm e}^{2\,z}}{c_{14}} \Big) u,\label{k3}\eny

\bny K^w_{~{w}}\(u,w,z\)=\frac18\,{u}^{2} \Big( {{\rm e}^{4\,z}}{u}^{2}{c_1}\,{w}^{4}+4\,{
{\rm e}^{4\,z}}{c_2}\,{u}^{2}{w}^{3}-4\,{{\rm e}^{4\,z}}{c_{15}}\,{u}^{2}{w}^{2}
-8\,{{\rm e}^{4\,z}}{c_{16}}\,{u}^{2}w\nonumber\\
+8\,u{
{\rm e}^{3\,z}}{c_4}\,{w}^{3}+8\,{{\rm e}^{4\,z}}{c_{20}}\,{u}
^{2}+12\,u{{\rm e}^{3\,z}}{w}^{2}{c_5}-2\,u{{\rm e}^{3\,z}}{w}^{2
}{c_6}\nonumber\\
-2\,{{\rm e}^{2\,z}}{u}^{2}{c_1}\,{w}^{2}-16\,u{
{\rm e}^{3\,z}}{c_{17}}\,w
+8\,{{\rm e}^{3\,z}}{c_7}\,uw-4\,{
{\rm e}^{2\,z}}{u}^{2}{c_2}\,w\nonumber\\+8\,u{{\rm e}^{3\,z}}{c_{18}}+8
\,u{{\rm e}^{3\,z}}{c_8}+4\,{{\rm e}^{2\,z}}{u}^{2}{c_{15}}+8
\,{{\rm e}^{2\,z}}{c_3}\,{u}^{2}\nonumber\\
+32\,{{\rm e}^{2\,z}}{c_9}\,
{w}^{2}-16\,{{\rm e}^{2\,z}}{c_{10}}\,w-8\,u{{\rm e}^{z}}{c_4}
\,w+8\,{{\rm e}^{2\,z}}{c_{11}}\nonumber\\+4\,{{\rm e}^{2\,z}}{c_{12}}+4\,u
{{\rm e}^{z}}{c_5}+2\,u{{\rm e}^{z}}{c_6}+{c_1}\,{u}^{2
} \Big),\nonumber\\\label{k4}\eny

\bny K^w_{~{z}}\(u,w,z\)=\frac18\,{u}^{2} \Big( -2\,{u}^{2}{{\rm e}^{2\,z}}{c_1}\,{w}^{3}+2\,
u{{\rm e}^{3\,z}}{w}^{4}{c_4}-6\,{u}^{2}{{\rm e}^{2\,z}}{w}^{2}{
c_2}+2\,u{{\rm e}^{3\,z}}{w}^{3}{c_5}\nonumber\\
-u{{\rm e}^{3\,z}}{w}^{
3}{c_6}-12\,u{{\rm e}^{z}}{c_4}\,{w}^{2}
+4\,{u}^{2}{{\rm e}^
{2\,z}}{c_{15}}\,w+16\,{{\rm e}^{2\,z}}{c_9}\,{w}^{3}-4\,u{
{\rm e}^{3\,z}}{c_{17}}\,{w}^{2}-6\,u{{\rm e}^{z}}w{c_5}\nonumber\\
+3\,u{
{\rm e}^{z}}w{c_6}-12\,{{\rm e}^{2\,z}}{c_{10}}\,{w}^{2}+4\,{u
}^{2}{{\rm e}^{2\,z}}{c_{16}}\nonumber\\
+4\,u{{\rm e}^{3\,z}}{c_{18}}\,w+2
\,{c_1}\,{u}^{2}w+2\,u{c_4}\,{{\rm e}^{-z}}+4\,u{{\rm e}^{z}
}{c_{17}}+4\,{{\rm e}^{2\,z}}{c_{12}}\,w+4\,u{{\rm e}^{3\,z}}{
c_{19}}\nonumber\\+2\,{c_2}\,{u}^{2}+4\,{{\rm e}^{2\,z}}{c_{13}}-16\,{
c_9}\,w+4\,{c_{10}} \Big),\nonumber\\\label{k5}\eny

\bny K^z_{~{z}}\(u,w,z\)=\frac12\, \Big( 2\,{{\rm e}^{2\,z}}{c_9}\,{w}^{4}-2\,{{\rm e}^{2\,z}
}{c_{10}}\,{w}^{3}-2\,u{{\rm e}^{z}}{c_4}\,{w}^{3}+{{\rm e}^{2
\,z}}{c_{12}}\,{w}^{2}+u{{\rm e}^{z}}{w}^{2}{c_6}\nonumber\\+{u}^{2}{c_1}\,{w}^{2}+2\,{{\rm e}^{-z}}{c_4}\,uw
+2\,{{\rm e}^{2\,z}}{
c_{13}}\,w+2\,u{{\rm e}^{z}}{c_7}\,w+2\,{u}^{2}{c_2}\,w+2
\,{c_9}\,{{\rm e}^{-2\,z}}+2\,{{\rm e}^{-z}}{c_5}\,u\nonumber\\
+2\,{
{\rm e}^{2\,z}}{c_{14}}+2\,u{{\rm e}^{z}}{c_8}+2\,{u}^{2}{c_3}-4\,{c_9}\,{w}^{2}+2\,{c_{10}}\,w+2\,{c_{11}} \Big) 
{u}^{2}.\nonumber\\\label{k6}\eny

To perform this analysis,  the field equations must admit contact
transformations which are  one-parameter transformations in the tangent bundle of the dynamical system.
In the next section, we give the form of the potential $V\(w,z\)$, the admitted contact Noether symmetry  and the quadratic
conservation laws which follow from the symmetry conditions (\ref{con1})-(\ref{con3}).

\section{Quadratic Conservation Laws}

We search for those interactions that admit  Noether contact symmetries. These symmetries ensure the existence of  conservation laws  that are linearly independent and in involution with the Hamiltonian. Stated differently,  we require that $\{I,H\}=0$ where $\{,\}$ is the Poisson bracket, and this  mandates that the  field equations form an integrable dynamical system. 

{ Furthermore, we emphasize that in \cite{p3}, linear conservation laws were found and also some cases where systems of linear equations were obtained. In theory, the latter  will   admit quadratic conservation laws. However, in that case the various potentials and solutions have already been reported using point symmetries. It is  unnecessary to use contact symmetries to re-derive those potentials. }

With the aid of (\ref{k1})-(\ref{k6}), we show that for
specific potential functions $V\(w,z\)$, where the effective potential is  defined as $V_{eff}=u^2V\(w,z\)$ , the Lagrangian (\ref{l0}) is
invariant under  contact symmetries.  For an arbitrary potential function the field
equations do not admit dynamical symmetries hence the field equations are not Noether integrable. 
In particular, we find that for the gravitational field equations (\ref{f11})-(\ref{f2}),
there are two specific choices of the potential function,  for which  quadratic in  momentum conservation laws
exist.

\subsubsection{Potential I}
In the case where $V(w,z)=V_0w^2e^{2z},$ the Lagrangian (\ref{l0}) possesses the dynamical symmetry
(\ref{con4}) which is explicitly of the form
\bn X^1=\(e^{2z}u^2w^2\dot{w}-2\,{\frac {  {{\rm e}^{2z_{{}}}} w_{{}}{u_{{}}}^{2}
}{ {{\rm e}^{2z_{{}}}}+1}}
\dot{z}\)\p_w+\(-2\,{\frac {  {{\rm e}^{2z_{{}}}} w_{{}}{z_{{}}}^{2}
}{  {{\rm e}^{2z_{{}}}} +1}}
\dot{w}+u^2\dot{z}\)\p_z,\label{x1}\en
and admits the quadratic conservation law
\bn
I^1= \left( {w_{{}}}^{2}{\dot{w}{{}}}^{2}{{\rm e}^{4\,z_{{}}}}-2\,w_{{}}\dot{w}{{}
}\dot{z}{{}}{{\rm e}^{2\,z_{{}}}}+{\dot{z}{{}}}^{2} \right) {u_{{}}}^{4}.\label{i1}\en

A second contact symmetry is

\bny X^2=\Big(\(\frac12\,{w_{{}}}^{4}{{\rm e}^{2\,z_{{}}}}+{w_{{}}}^{2}+\frac12\,{{\rm e}^{-2
\,z_{{}}}}\)
\dot{u}-2\,{\frac {u_{{}}w_{{}} \left( {w_{{}}}^{2}{{\rm e}^{2\,z_{{}}}}+1
 \right) }{{{\rm e}^{2\,z_{{}}}}{u_{{}}}^{2}-1}}
\dot{w}\nonumber\\-{\frac {{{\rm e}^{-z_{{}}}} \left( {{\rm e}^{5\,z_{{}}}}{u_{{}}}^{2}{
w_{{}}}^{4}-{{\rm e}^{3\,z_{{}}}}{w_{{}}}^{4}-{{\rm e}^{z_{{}}}}{u_{{}
}}^{2}+{{\rm e}^{-z_{{}}}} \right) u_{{}}}{ \left( {{\rm e}^{2\,z_{{}}
}}{u_{{}}}^{2}-1 \right)  \left( {u_{{}}}^{2}-1 \right) }}
\dot{z}\Big)\p_u\nonumber\\
-\(2\,{\frac {u_{{}}w_{{}} \left( {w_{{}}}^{2}{{\rm e}^{2\,z_{{}}}}+1
 \right) }{{{\rm e}^{2\,z_{{}}}}{u_{{}}}^{2}-1}}
\dot{u}+2w^2\dot{w}+2\,{\frac {w_{{}} \left( {w_{{}}}^{2}{{\rm e}^{2\,z_{{}}}}-1 \right) 
}{{{\rm e}^{2\,z_{{}}}}+1}}
\dot{z}\)\p_w-\nonumber\\
\Big({\frac {{{\rm e}^{-z_{{}}}} \left( {{\rm e}^{5\,z_{{}}}}{u_{{}}}^{2}{
w_{{}}}^{4}-{{\rm e}^{3\,z_{{}}}}{w_{{}}}^{4}-{{\rm e}^{z_{{}}}}{u_{{}
}}^{2}+{{\rm e}^{-z_{{}}}} \right) u_{{}}}{ \left( {{\rm e}^{2\,z_{{}}
}}{u_{{}}}^{2}-1 \right)  \left( {u_{{}}}^{2}-1 \right) }}
\dot{u}+2\,{\frac {w_{{}} \left( {w_{{}}}^{2}{{\rm e}^{2\,z_{{}}}}-1 \right) 
}{{{\rm e}^{2\,z_{{}}}}+1}}
\dot{w}\nonumber\\-\(\frac12\, \left( -{w_{{}}}^{4}{{\rm e}^{4\,z_{{}}}}+2\,{w_{{}}}^{2}{
{\rm e}^{2\,z_{{}}}}-1 \right) {{\rm e}^{-2\,z_{{}}}}
\)\dot{z}\Big)\p_z,\label{x2}\eny
and contact Noether integral

\bny I^2=\frac12\,\frac{1}{{{{\rm e}^
{2\,z_{{}}}}{u}^{2}-1}} \Big(4\,{{\rm e}^{2\,z_{{}}}}{u}^{2}{w}^{2}{\dot{w}}^{2}-4\,{
{\rm e}^{4\,z_{{}}}}{u}^{4}{w}^{2}{\dot{w}}^{2}+2\,{{\rm e}^{2\,z_{{}}
}}{u}^{4}{w}^{2}{\dot{z}}^{2}-{{\rm e}^{4\,z_{{}}}}{u}^{4}{w}^{4}{\dot{z}{{
}}}^{2}\nonumber\\
+{{\rm e}^{2\,z_{{}}}}{u}^{2}{w}^{4}{\dot{z}}^{2}+4\,u\dot{u}w
\dot{w}-4\,{u}^{2}w\dot{w}\dot{z}-{{\rm e}^{4\,z_{{}}}}{u}^{2}{\dot{u}
}^{2}{w}^{4}-2\,{{\rm e}^{2\,z_{{}}}}{u}^{2}{\dot{u}}^{2}{w}^{2}-2\,{
{\rm e}^{-2\,z_{{}}}}u\dot{u}\dot{z}\nonumber\\
-4\,{{\rm e}^{2\,z_{{}}}}{u}^{3}\dot{u}
{{}}w\dot{w}
+4\,{{\rm e}^{2\,z_{{}}}}u\dot{u}{w}^{3}\dot{w}+2\,{
{\rm e}^{2\,z_{{}}}}u\dot{u}{w}^{4}\dot{z}-2\,{{\rm e}^{4\,z_{{}}}}{u}
^{3}\dot{u}{w}^{4}\dot{z}-4\,{{\rm e}^{4\,z_{{}}}}{u}^{4}{w}^{3}\dot{w}{{}
}\dot{z}\nonumber\\
-4\,{{\rm e}^{4\,z_{{}}}}{u}^{3}\dot{u}{w}^{3}\dot{w}+4\,{{\rm e}^{2\,z_{{}}}}{u}^{4}w\dot{w}\dot{z}
+4\,{{\rm e}^{2\,z_{{}}}}{u}
^{2}{w}^{3}\dot{w}\dot{z}+{{\rm e}^{2\,z_{{}}}}{\dot{u}}^{2}{w}^{4}+2
\,{u}^{3}\dot{u}\dot{z}\nonumber\\-2\,{u}^{2}{w}^{2}{\dot{z}}^{2}+{{\rm e}^{-2\,z
_{{}}}}{u}^{2}{\dot{z}}^{2}-{u}^{4}{\dot{z}}^{2}-{u}^{2}{\dot{u}}^{2}+
2\,{\dot{u}}^{2}{w}^{2}+{{\rm e}^{-2\,z_{{}}}}{\dot{u}}^{2}\Big).\nonumber\\ \label{i2}
\eny
When the contact symmetry (\ref{con4}) is
\bny X^3=\(\frac12\,{w_{{}}}^{2}{{\rm e}^{2\,z_{{}}}}\dot{u}-{\frac {u_{{}}w_{{}}{{\rm e}^{2\,z_{{}}}}}{{{\rm e}^{2\,z_{{}}}}{u_{{
}}}^{2}-1}}
\dot{w}-{\frac {{{\rm e}^{2\,z_{{}}}}u_{{}}{w_{{}}}^{2}}{{u_{{}}}^{2}-1}}\dot{z}\)\p_u\nonumber\\+\(-{\frac {u_{{}}w_{{}}{{\rm e}^{2\,z_{{}}}}}{{{\rm e}^{2\,z_{{}}}}{u_{{
}}}^{2}-1}}
\dot{u}-\frac12\dot{w}-{\frac {w_{{}}{{\rm e}^{2\,z_{{}}}}}{{{\rm e}^{2\,z_{{}}}}+1}}\dot{z}\)\p_w\nonumber\\
+\(-{\frac {{{\rm e}^{2\,z_{{}}}}u_{{}}{w_{{}}}^{2}}{{u_{{}}}^{2}-1}}\dot{u}-{\frac {w_{{}}{{\rm e}^{2\,z_{{}}}}}{{{\rm e}^{2\,z_{{}}}}+1}}\dot{w}-\frac12\,{w_{{}}}^{2}  {{\rm e}^{2z_{{}}}} \dot{z}\)\p_z,\label{x3}\eny

we obtain the quadratic  integral
\bny I^3=\frac12\, \left( -{u}^{2}{w}^{2}{\dot{z}}^{2}+2\,V_0{u}^{2}{w}^{2}-2\,{u}^{2
}w\dot{w}\dot{z}-2\,u{w}^{2}\dot{u}\dot{z}-{u}^{2}{\dot{w}}^{2}-2\,uw\dot{u}
{{}}\dot{w}-{w}^{2}{\dot{u}}^{2} \right) {{\rm e}^{2\,z_{{}}}},\nonumber\\\label{i3}\eny
The Lagrangian admits the symmetry
\bn X^4=\(e^{2z}\dot{u}-2\,{\frac {{{\rm e}^{2\,z_{{}}}}u_{{}}}{{u_{{}}}^{2}-1}}\dot{z}\)\p_u+\(-2\,{\frac {{{\rm e}^{2\,z_{{}}}}u_{{}}}{{u_{{}}}^{2}-1}}\dot{u}-e^{2z}\dot{z}\)\p_z,\label{x4}\en

with conservation law
\bn I^4= \left( -{u}^{2}{\dot{z}}^{2}-2\,u\dot{u}\dot{z}-{\dot{u}}
^{2} \right) {{\rm e}^{2\,z_{{}}}}.\label{i4}\en

Lastly, the following two dynamical symmetries 
\bny X^5=\({\frac {{u_{{}}}^{2}w_{{}}{{\rm e}^{z_{{}}}} \left( {w_{{}}}^{2}{
{\rm e}^{2\,z_{{}}}}+1 \right) }{{{\rm e}^{2\,z_{{}}}}{u_{{}}}^{2}-1}}
\dot{w}-{\frac {{{\rm e}^{-z_{{}}}} \left( {w_{{}}}^{2}{{\rm e}^{2\,z_{{}}}}+
1 \right) {u_{{}}}^{2}}{{u_{{}}}^{2}-1}}
\dot{z}\)\p_u\nonumber\\
+\({\frac {{u_{{}}}^{2}w_{{}}{{\rm e}^{z_{{}}}} \left( {w_{{}}}^{2}{
{\rm e}^{2\,z_{{}}}}+1 \right) }{{{\rm e}^{2\,z_{{}}}}{u_{{}}}^{2}-1}}
\dot{u}+2\,{{\rm e}^{z_{{}}}}u_{{}}{w_{{}}}^{2}\dot{w}+{\frac {w_{{}} \left( {w_{{}}}^{2}{{\rm e}^{2\,z_{{}}}}-3 \right) {
{\rm e}^{z_{{}}}}u_{{}}}{{{\rm e}^{2\,z_{{}}}}+1}}
\dot{z}\)\p_w\nonumber\\
+\(-{\frac {{{\rm e}^{-z_{{}}}} \left( {w_{{}}}^{2}{{\rm e}^{2\,z_{{}}}}+
1 \right) {u_{{}}}^{2}}{{u_{{}}}^{2}-1}}
\dot{u}+{\frac {w_{{}} \left( {w_{{}}}^{2}{{\rm e}^{2\,z_{{}}}}-3 \right) {
{\rm e}^{z_{{}}}}u_{{}}}{{{\rm e}^{2\,z_{{}}}}+1}}
\dot{w}-{{\rm e}^{-z_{{}}}}u_{{}} \left( {w_{{}}}^{2}{{\rm e}^{2\,z_{{}}}}-1
 \right) 
\dot{z}\)\p_z,\nonumber\\\label{x5}\eny

\bny X^6=\({\frac {{u_{{}}}^{2}w_{{}}{{\rm e}^{3\,z_{{}}}}}{{u_{{}}}^{2}{{\rm e}^
{2\,z_{{}}}}-1}}
\dot{w}-{\frac {{{\rm e}^{z_{{}}}}{u_{{}}}^{2}}{{u_{{}}}^{2}-1}}\dot{z}\)\p_u+\({\frac {{u_{{}}}^{2}w_{{}}{{\rm e}^{3\,z_{{}}}}}{{u_{{}}}^{2}{{\rm e}^
{2\,z_{{}}}}-1}}
\dot{u}+{\frac {w_{{}}u_{{}}{{\rm e}^{3\,z_{{}}}}}{{{\rm e}^{2\,z_{{}}}}+1}}\dot{z}\)\p_w\nonumber\\
+\(-{\frac {{{\rm e}^{z_{{}}}}{u_{{}}}^{2}}{{u_{{}}}^{2}-1}}\dot{u}+{\frac {w_{{}}u_{{}}{{\rm e}^{3\,z_{{}}}}}{{{\rm e}^{2\,z_{{}}}}+1}}\dot{w}-e^{z}u\dot{z}\)\p_z,\label{x6}\eny
admit the next two quadratic conservation laws of the field equations,

\bny I^5=2\,{{\rm e}^{3\,z_{{}}}}{u}^{3}{w}^{2}{\dot{w}}^{2}-{{\rm e}^{z_{{}}}}
{u}^{3}{w}^{2}{\dot{z}}^{2}+{{\rm e}^{3\,z_{{}}}}{u}^{3}{w}^{3}\dot{w}
\dot{z}-{{\rm e}^{z_{{}}}}{u}^{2}\dot{u}{w}^{2}\dot{z}\nonumber\\+{{\rm e}^{3\,z_
{{}}}}{u}^{2}\dot{u}{w}^{3}\dot{w}+{{\rm e}^{-z_{{}}}}{u}^{3}{\dot{z}}
^{2}-3\,{{\rm e}^{z_{{}}}}{u}^{3}w\dot{w}\dot{z}-{{\rm e}^{-z_{{}}}}{u
}^{2}\dot{u}\dot{z}\nonumber\\
+{{\rm e}^{z_{{}}}}{u}^{2}\dot{u}w\dot{w},\label{i5}\eny


\bn I^6= \left( {{\rm e}^{2\,z_{{}}}}uw\dot{w}\dot{z}+{{\rm e}^{2\,z_{{}}}}\dot{u}{
{}}w\dot{w}-u{\dot{z}}^{2}-\dot{u}\dot{z} \right) {u}^{2}{{\rm e}^{z_
{{}}}}.\label{i6}\en
respectively. As we have seen, this case produces six quadratic conservation laws.

\subsubsection{Potential II}
In the second case, let $V(w,z)=V_0$.
This case admits $X^1$ and $X^3$ and hence $I^1$ and $I^3$. It also admits $X^4$, but a difference in the boundary term provides $\bar{I}^4=I^4+2\,V_0{u}^{2}{{\rm e}^{2\,z_{{}}}}.$ We list the remaining contact symmetries and integrals below.

For the dynamical symmetry
\bny X^7=\(\({w_{{}}}^{4}{{\rm e}^{2\,z_{{}}}}+2\,{w_{{}}}^{2}+{{\rm e}^{-2\,z_{{}}
}}\)
\dot{u}-4\,{\frac {w_{{}} \left( {w_{{}}}^{2}{{\rm e}^{2\,z_{{}}}}+1 \right) 
u_{{}}}{{{\rm e}^{2\,z_{{}}}}{u_{{}}}^{2}-1}}
\dot{w}-2\,{\frac {u_{{}} \left( {w_{{}}}^{4}{{\rm e}^{2\,z_{{}}}}-{{\rm e}^{
-2\,z_{{}}}} \right) }{{u_{{}}}^{2}-1}}
\dot{z}\)\p_u\nonumber\\
+\(-4\,{\frac {w_{{}} \left( {w_{{}}}^{2}{{\rm e}^{2\,z_{{}}}}+1 \right) 
u_{{}}}{{{\rm e}^{2\,z_{{}}}}{u_{{}}}^{2}-1}}
\dot{u}-4w^2\dot{w}-4\,{\frac {w_{{}} \left( {w_{{}}}^{2}{{\rm e}^{2\,z_{{}}}}-1 \right) 
}{{{\rm e}^{2\,z_{{}}}}+1}}
\dot{z}\)\p_w\nonumber\\+\(-2\,{\frac {u_{{}} \left( {w_{{}}}^{4}{{\rm e}^{2\,z_{{}}}}-{{\rm e}^{
-2\,z_{{}}}} \right) }{{u_{{}}}^{2}-1}}
\dot{u}-4\,{\frac {w_{{}} \left( {w_{{}}}^{2}{{\rm e}^{2\,z_{{}}}}-1 \right) 
}{{{\rm e}^{2\,z_{{}}}}+1}}
\dot{w}+\(-{w_{{}}}^{4}{{\rm e}^{2\,z_{{}}}}-{{\rm e}^{-2\,z_{{}}}}+2\,{w_{{}}}^
{2}
\)\dot{z}\)\p_z,\nonumber\\\label{x7}\eny

we determine that the quadratic integral (\ref{i0}) is of the form
\bny I^7=-{{\rm e}^{2\,z_{{}}}}{u}^{2}{w}^{4}{\dot{z}}^{2}+2\,{w}^{4}V_0{u}^{2}{
{\rm e}^{2\,z_{{}}}}-4\,{{\rm e}^{2\,z_{{}}}}{u}^{2}{w}^{3}\dot{w}\dot{z}{{
}}-2\,{{\rm e}^{2\,z_{{}}}}u\dot{u}{w}^{4}\dot{z}\nonumber\\
-4\,{{\rm e}^{2\,z_{
{}}}}{u}^{2}{w}^{2}{\dot{w}}^{2}-4\,{{\rm e}^{2\,z_{{}}}}u\dot{u}{w}^{
3}\dot{w}-{{\rm e}^{2\,z_{{}}}}{\dot{u}}^{2}{w}^{4}+2\,{u}^{2}{w}^{2}{
\dot{z}}^{2}-{{\rm e}^{-2\,z_{{}}}}{u}^{2}{\dot{z}}^{2}\nonumber\\
+4\,{w}^{2}V_0{u}
^{2}+4\,{u}^{2}w\dot{w}\dot{z}+2\,{u}^{2}V_0{{\rm e}^{-2\,z_{{}}}}+2\,{
{\rm e}^{-2\,z_{{}}}}u\dot{u}\dot{z}-4\,u\dot{u}w\dot{w}\nonumber\\
-2\,{\dot{u}}^
{2}{w}^{2}-{{\rm e}^{-2\,z_{{}}}}{\dot{u}}^{2}.\label{i7}\eny
If the generator is
\bny X^8=\(\({w_{{}}}^{3}{{\rm e}^{2\,z_{{}}}}+w_{{}}\)\dot{u}+{\frac { \left( 3\,{w_{{}}}^{2}{{\rm e}^{2\,z_{{}}}}+1 \right) u_{{}}
}{-  {{\rm e}^{2z_{{}}}} {u_{{}}}^{2}+1}}
\dot{w}-2\,{\frac {u_{{}}{w_{{}}}^{3}{{\rm e}^{2\,z_{{}}}}}{{u_{{}}}^{2}-1}}\dot{z}\)\p_u\nonumber\\
+\({\frac { \left( 3\,{w_{{}}}^{2}{{\rm e}^{2\,z_{{}}}}+1 \right) u_{{}}
}{- {{\rm e}^{2z_{{}}}} {u_{{}}}^{2}+1}}
\dot{u}-2w\dot{w}-{\frac {3\,{w_{{}}}^{2}{{\rm e}^{2\,z_{{}}}}-1}{{{\rm e}^{2\,z_{{}}}}
+1}}\dot{z}
\)\p_w\nonumber\\
+\(-2\,{\frac {u_{{}}{w_{{}}}^{3}{{\rm e}^{2\,z_{{}}}}}{{u_{{}}}^{2}-1}}\dot{u}-{\frac {3\,{w_{{}}}^{2}{{\rm e}^{2\,z_{{}}}}-1}{{{\rm e}^{2\,z_{{}}}}
+1}}
\dot{w}+\(-{w_{{}}}^{3}{{\rm e}^{2\,z_{{}}}}+w_{{}}\)\dot{z}\)\p_z,\label{x8}\eny
then the integral is

\bny I^8=-{{\rm e}^{2\,z_{{}}}}{u}^{2}{w}^{3}{\dot{z}}^{2}+2\,{w}^{3}V_0{u}^{2}{
{\rm e}^{2\,z_{{}}}}-3\,{{\rm e}^{2\,z_{{}}}}{u}^{2}{w}^{2}\dot{w}\dot{z}{{
}}\nonumber\\
-2\,{{\rm e}^{2\,z_{{}}}}u\dot{u}{w}^{3}\dot{z}-2\,{{\rm e}^{2\,z_{
{}}}}{u}^{2}w{\dot{w}}^{2}-3\,{{\rm e}^{2\,z_{{}}}}u\dot{u}{w}^{2}\dot{w}{{
}}-{{\rm e}^{2\,z_{{}}}}{\dot{u}}^{2}{w}^{3}+{u}^{2}w{\dot{z}}^{2}\nonumber\\
+2
\,wV_0{u}^{2}+{u}^{2}\dot{w}\dot{z}-u\dot{u}\dot{w}-{\dot{u}}^{2}w.\label{i8}\eny

The Lagrangian (\ref{l0}), also admits the contact symmetry
\bny X^9=\(w_{{}}{{\rm e}^{2\,z_{{}}}}\dot{u}+{\frac {{{\rm e}^{2\,z_{{}}}}u_{{}}}{-{{\rm e}^{2\,z_{{}}}}{u_{{}}}^{2
}+1}}
\dot{w}-2\,{\frac {w_{{}}{{\rm e}^{2\,z_{{}}}}u_{{}}}{{u_{{}}}^{2}-1}}\dot{z}\)\p_u \nonumber\\+\({\frac {{{\rm e}^{2\,z_{{}}}}u_{{}}}{-{{\rm e}^{2\,z_{{}}}}{u_{{}}}^{2
}+1}}
\dot{u}-{\frac {{{\rm e}^{2\,z_{{}}}}}{{{\rm e}^{2\,z_{{}}}}+1}}\dot{z}\)\p_w\nonumber\\
+\(-2\,{\frac {w_{{}}{{\rm e}^{2\,z_{{}}}}u_{{}}}{{u_{{}}}^{2}-1}}\dot{u}-{\frac {{{\rm e}^{2\,z_{{}}}}}{{{\rm e}^{2\,z_{{}}}}+1}}\dot{w}-w_{{}}{{\rm e}^{2\,z_{{}}}}\dot{z}\)\p_z,\label{x9}\eny

and Noether integral
\bn I^9= \left( -{u}^{2}w{\dot{z}}^{2}+2\,V_0{u}^{2}w-{u}^{2}\dot{w}\dot{z}-2\,u
w\dot{u}\dot{z}-u\dot{u}\dot{w}-w{\dot{u}}^{2} \right) {{\rm e}^{2\,z_
{{}}}}.\label{i9}\en



Continuing with our list, we have four more dynamical symmetries
\bny X^{10}=\(\(\frac18\,{u_{{}}}^{2} \left( {w_{{}}}^{4}{{\rm e}^{2\,z_{{}}}}-2\,{w_{{}}}
^{2}+{{\rm e}^{-2\,z_{{}}}} \right)
\)\dot{w}-\frac12\,{\frac {{u_{{}}}^{2}w_{{}} \left( {w_{{}}}^{2}{{\rm e}^{2\,z_{{}
}}}-1 \right) }{{{\rm e}^{2\,z_{{}}}}+1}}
\dot{z}\)\p_w\nonumber\\
+\(-\frac12\,{\frac {{u_{{}}}^{2}w_{{}} \left( {w_{{}}}^{2}{{\rm e}^{2\,z_{{}
}}}-1 \right) }{{{\rm e}^{2\,z_{{}}}}+1}}
\dot{w}+\frac12\,{w_{{}}}^{2}{u_{{}}}^{2}\dot{z}\)\p_z,\label{x10}\eny

\bny X^{11}=\(\frac12\,{u_{{}}}^{2}w_{{}} \left( {w_{{}}}^{2}{{\rm e}^{2\,z_{{}}}}-1
 \right)
\dot{w}-\frac12\,{\frac {{u_{{}}}^{2} \left( 3\,{w_{{}}}^{2}{{\rm e}^{2\,z_{{}}}}
-1 \right) }{{{\rm e}^{2\,z_{{}}}}+1}}
\dot{z}\)\p_w\nonumber\\
+\(-\frac12\,{\frac {{u_{{}}}^{2} \left( 3\,{w_{{}}}^{2}{{\rm e}^{2\,z_{{}}}}
-1 \right) }{{{\rm e}^{2\,z_{{}}}}+1}}
\dot{w}+u^2w\dot{z}\)\p_z,\label{x11}\eny

\bny X^{12}=\(-{u_{{}}}^{2} \left( {w_{{}}}^{2}{{\rm e}^{2\,z_{{}}}}-1 \right) \dot{w}+2\,{\frac {w_{{}}{{\rm e}^{2\,z_{{}}}}{u_{{}}}^{2}}{{{\rm e}^{2\,z_{{
}}}}+1}}
\dot{z}\)\p_w+\(2\,{\frac {w_{{}}{{\rm e}^{2\,z_{{}}}}{u_{{}}}^{2}}{{{\rm e}^{2\,z_{{
}}}}+1}}
\dot{w}\)\p_z,\label{x12}\eny

\bn X^{13}=\(\frac12\,{u_{{}}}^{2}w_{{}}{{\rm e}^{2\,z_{{}}}}\dot{w}-\frac12\,{\frac {{{\rm e}^{2\,z_{{}}}}{u_{{}}}^{2}}{{{\rm e}^{2\,z_{{}}}}
+1}}
\dot{z}\)\p_w+\(-\frac12\,{\frac {{{\rm e}^{2\,z_{{}}}}{u_{{}}}^{2}}{{{\rm e}^{2\,z_{{}}}}
+1}}
\dot{w}\)\p_z.\label{x13}\en

Consequently, the associated Noether contact integrals are, respectively,
\bn I^{10}=\frac18\, \left( {{\rm e}^{4\,z_{{}}}}{w}^{4}{\dot{w}}^{2}-4\,{{\rm e}^{2
\,z_{{}}}}{w}^{3}\dot{w}\dot{z}-2\,{{\rm e}^{2\,z_{{}}}}{w}^{2}{\dot{w}{{}
}}^{2}+{\dot{w}}^{2}+4\,{w}^{2}{\dot{z}}^{2}+4\,w\dot{w}\dot{z}
 \right) {u}^{4},\label{i10}\en


\bn I^{11}=-\frac12\, \left( -{{\rm e}^{4\,z_{{}}}}{w}^{3}{\dot{w}}^{2}+3\,{{\rm e}^{
2\,z_{{}}}}{w}^{2}\dot{w}\dot{z}+{{\rm e}^{2\,z_{{}}}}w{\dot{w}}^{2}-2
\,w{\dot{z}}^{2}-\dot{w}\dot{z} \right) {u}^{4},\label{i11}\en


\bn I^{12}=- \left( {w}^{2}\dot{w}{{\rm e}^{2\,z_{{}}}}-2\,w\dot{z}-\dot{w}
 \right) {u}^{4}{{\rm e}^{2\,z_{{}}}}\dot{w},\label{i12}\en

and finally,


\bn I^{13}=\frac12\, \left( w\dot{w}{{\rm e}^{2\,z_{{}}}}-\dot{z} \right) {u}^{4}{
{\rm e}^{2\,z_{{}}}}\dot{w}.\label{i13}\en

The next natural step is to find 
analytical solutions for the two different potentials.

\section{Analytical Solutions}

In order to proceed with the solution of the system of equations corresponding to each of the two potential functions, some manipulation is necessary.   
We then write the Euler-Lagrange equations for each potential and solve them taking into account the constraint imposed by the associated Hamiltonian.
To proceed, we first apply the transformations \cite{p3}, {which give the normal coordinates and are found using the invariant functions associated with the Killing vectors:}
\bny u^2={x}^2-{y}^2-s^2,\nonumber\\
w=\frac{s}{{x}+{y}},\nonumber\\
z=\ln \frac{{x}+{y}}{\sqrt{{x}^2-{y}^2-s^2}}. \label{t1}\eny

In the new coordinate system $\{x,y,s\}$ the Lagrangian  (\ref{l0}) takes on the simpler form
\bn L\(x,\dot{x},y,\dot{y},s,\dot{s}\)=-\frac12\dot{x}^2+\frac12\dot{y}^2+\frac12\dot{s}^2-\({x}^2-{y}^2-s^2\) V\(x,y,s\).\label{l1}\en

\begin{figure}[!ht]%
\centering
\subfigure[][]{%
\label{f1}%
\includegraphics[height=2in]{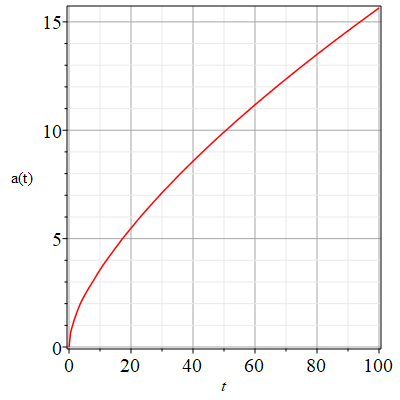}}%
\hspace{8pt}%
\subfigure[][]{%
\label{f4}%
\includegraphics[height=2in]{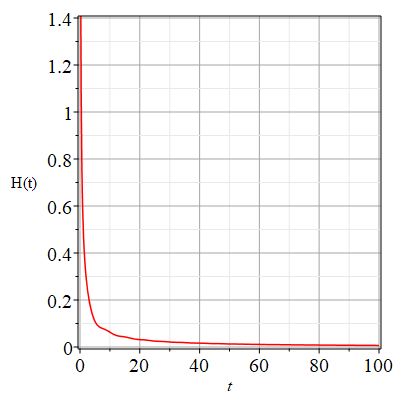}}\\
\subfigure[][]{%
\label{f5}%
\includegraphics[height=2in]{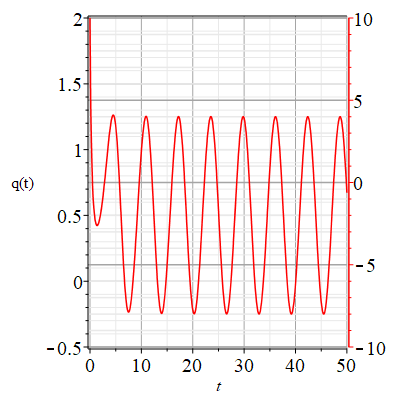}}%
\subfigure[][]{%
\label{f6}%
\includegraphics[height=2in]{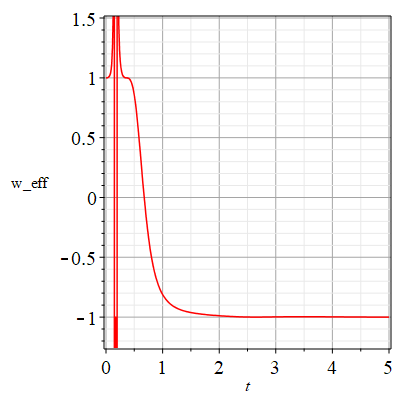}}
\caption[A set of four subfigures.]{Evolution of the scale factor (\ref{a2}), the corresponding Hubble function $H(t)$, the deceleration parameter $q(t)$ and  the  equation of state parameter 
$w_{eff}.$ Parameter values are :
\subref{f1}  $s_0=1,~\bar{V}_0=0.5$; 
\subref{f4} $\bar{V}_0=0.470$; 
\subref{f5} $s_0=1,~\bar{V}_0=0.5$; and
\subref{f6} $\bar{V}_0=1,~s_0=\sqrt{3},~x_0=2,~y_0=1$.}%
\label{fig6}%
\end{figure}

In the case of Potential I, 
and if we
let $\bar{V}_0=\sqrt {2 V_0
}$, the analytical solutions of the dynamical system associated with (\ref{l1}) are
\bny s \left( t \right) &=&{s_0}\,\sin \left( \bar{V}_0 t \right) +{s_1}\,\cos \left( \bar{V}_0 t \right) ,\nonumber\\
x
 \left( t \right) &=&{x_0}\,t+{x_1},\nonumber\\
 y \left( t \right)&=&{y_0}\,t+{y_1}, \eny
 with Hamiltonian constraint $E=s_1^{2}V_0+V_0s_0^{2}+\frac12\,y_0^{2}-\frac12\,x_0^{2}
.$ One can easily transform back into $\{u,w,z\}$ if necessary.
The scale factor here is 
\bn a^3(t)=\frac38\, \Big(\left( {x_0}\,t+{x_1} \right) ^{2}-\, \left( {y_0}\,t+{y_1} \right) ^{2}-\, \left( {s_0}\,\sin \left( 
 \bar{V}_0 t \right) +{s_1}\,\cos \left(  \bar{V}_0 t \right)  \right) ^{2}\Big),
 \label{a25}\en and using the singularity condition $a\(t\to0\)=0,$ we have the constraint for the constants of integration viz. $\frac38 \(x_1^2-y_1^2-s_1^2\)=0.$ { If we set $s_0=s_1, x_0=x_1, y_0=y_1,$ the scale factor (\ref{a25}) reduces to
 \bn a^3(t)=\frac38 s_0^2\, \Big(t^2+2t-\sin(2 \bar{V}_0 t)\Big).
 \label{a2}\en In turn, the Hubble function (defined as $H=\frac{\dot{a}}{a}$) is 
 \bn H(t)={\frac {2\,\bar{V}_0\cos \left( 2\,\bar{V}_0t \right) -2\,t-2}{-3\,{t}^{2}+3\,\sin
 \left( 2\,\bar{V}_0t \right) -6\,t}},\label{h11}\en
 and the deceleration parameter is
 \bny q(t)&=& {\frac {-2\, \left( \cos \left( 2\,\bar{V_0}t \right)  \right) ^{2}{\bar{V_0}}^{2
}-8\,\bar{V_0} \left( t+1 \right) \cos \left( 2\,\bar{V_0}t \right) + }{ 2\left( -\bar{V_0}\cos \left( 2\,\bar{V_0}t
 \right) +t+1 \right) ^{2}}}\nonumber\\
 &&+\frac12\frac{\left( 3+
 \left( -6\,{t}^{2}-12\,t \right) {\bar{V_0}}^{2} \right) \sin \left( 2\,\bar{V_0}t
 \right) +{t}^{2}+6\,{\bar{V_0}}^{2}+2\,t+4}{ \left( -\bar{V_0}\cos \left( 2\,\bar{V_0}t
 \right) +t+1 \right) ^{2}}.\nonumber\\\label{qq}\eny

In Figure \ref{fig6}, we explore the general dynamical behaviour of this model. It is clear that $H(t)$ has the
largest value at early times and then suddenly declines. The deceleration parameter $q(t)$ starts off with a positive value and then begins to oscillate, $-0.249\leq q\leq 2$. Oscillating cosmological models have appeared in the literature \cite{osc1,osc2,osc3} although different from our model, some showcased 
the cosmic transit phenomenon signified by  a so-called 
flipping behaviour
of the deceleration parameter. These models may be important for the coincidence
problem due to periods of acceleration \cite{osc4}. Here, we observe that $q$ varies periodically between mostly a
 decelerated phase ($q$ positive) and an accelerating phase ($q$ negative). As for the equation of state parameter 
$w_{eff},$ after some initial volatility,  we find $w_{eff}\to-1$, predicting an equation of state 
 behaviour  similar to  $\Lambda$-cosmology ($\Lambda$ Cold Dark Matter) model.}
 
On the other hand, in terms of Potential II, we have a more interesting solution, viz. 
{
\bny s \left( t \right) &=&s_0\exp\(\bar{V_0}~t\)+s_1\exp\(-\bar{V_0}~t\),\nonumber\\
x \left( t \right) &=&x_0\exp\(\bar{V_0}~t\)+x_1\exp\(-\bar{V_0}~t\),\nonumber\\
y \left( t \right) &=&y_0\exp\(\bar{V_0}~t\)+y_1\exp\(-\bar{V_0}~t\), \eny
 with Hamiltonian constraint $E=-4\, \left( y_0y_1-x_0x_1+s_0s_1 \right) V_0
.$
The scale factor  becomes
\bn a^3(t)=\frac38\Big(  \left( {x_0}\,{{\rm e}^{\bar{V_0}t}}+{x_1}\,{
{\rm e}^{-\bar{V_0}t}} \right) ^{2}- \left( {y_0}\,{
{\rm e}^{\bar{V_0}t}}+{y_1}\,{{\rm e}^{-\bar{V_0}t}} \right) ^{2}- \left( {s_0}\,{{\rm e}^{\bar{V_0}t}}
+{s_1}\,{{\rm e}^{-\bar{V_0}t}} \right) ^{2}
\Big)
, \label{nov1}\en
and  the singularity condition $a\(t\to0\)=0,$ gives the added constraint
$-{{y_0}}^{2}-{{y_1}}^{2}+{{x_0}}^{2}+{{x_1}}^{2}-{{s_0}}^{2}
-{{s_1}}^{2}+\frac{E}{2V_0}=0.
$

The Hubble parameter in this case is
\bn H(t)={\frac {2 \left( {{\rm e}^{4\,\bar{V_0}t}} \left( {{y_0}}^{2}-{{x_0}}^{2}+{{s_0}}^{2} \right) -{{y_1}}^{2}
+{{x_1}}^{2}-{{s_1}}^{2} \right) \bar{V_0}}{
 \left( 3\,{{y_0}}^{2}-3\,{{x_0}}^{2}+3\,{{s_0}}^{2}
 \right) {{\rm e}^{4\,\bar{V_0}t}}+ \left( 6\,{y_0}\,{
y_1}-6\,{x_0}\,{x_1}+6\,{s_0}\,{s_1} \right) 
{{\rm e}^{2\,\bar{V_0}t}}+3\,{{y_1}}^{2}-3\,{{x_1}}
^{2}+3\,{{s_1}}^{2}}}, \label{nov2}
\en and the decelation parameter
\bny q(t)=\frac {-10\, \left( {{y_1}}^{2}-{{x_1}}^{2}+{{s_1}}^{2
} \right)  \left( {{y_0}}^{2}-{{x_0}}^{2}+{{s_0}}^{2}
 \right) {{\rm e}^{4\,\bar{V_0}t}}+6\, \left( {{y_1}}^{2
}-{{x_1}}^{2}+{{s_1}}^{2} \right) \frac{E}{4V_0} {{\rm e}^{
2\,\bar{V_0}t}}}{ \left(  \left( {{
y_0}}^{2}-{{x_0}}^{2}+{{s_0}}^{2} \right) {{\rm e}^{4\,
\bar{V_0}t}}-{{y_1}}^{2}+{{x_1}}^{2}-{{s_1}}^{
2} \right) ^{2}}\nonumber\\+\frac{6\, \left( {{y_0}}^{2}-{{x_0}}^{2}+
{{s_0}}^{2} \right) \frac{E}{4V_0} {{\rm e}^{6\,\bar{V_0}t}}- \left( {{y_0}}^{2}-{{x_0}}^{2}+{{s_0}}^{2}
 \right) ^{2}{{\rm e}^{8\,\bar{V_0}t}}- \left( {{y_1}}^{
2}-{{x_1}}^{2}+{{s_1}}^{2} \right) ^{2}}{ \left(  \left( {{
y_0}}^{2}-{{x_0}}^{2}+{{s_0}}^{2} \right) {{\rm e}^{4\,
\bar{V_0}t}}-{{y_1}}^{2}+{{x_1}}^{2}-{{s_1}}^{
2} \right) ^{2}}.\nonumber\\\label{nov3}
\eny

 A standard calculation also reveals that the equation of state parameter is
$w_{eff}=-1.$ A clear  resemblance to $\Lambda$-cosmology can  be seen in the special case of $x_0=-x_1,~y_0=-y_1,~ s_0=-s_1,$ where 
the scale factor  becomes}
\bn a^3(t)=\frac38\, \frac{E}{4V_0}\sinh^2\(\bar{V_0}~t\). \label{aa1}\en
 Recall that the exact solution of the scale factor of $\Lambda$-cosmology is $$a^3_{\Lambda}(t)=\(\frac{\Omega_{m0}}{1-\Omega_{m0}}\)\sinh^2\(k~t\),$$ where $k=\frac32 H_0\sqrt{1-\Omega_{m0}}$. 
 {This special case identifies the scalar fields $z$ and $w$ to be constant, that is $w(t)=w_0$ and $z(t)=z_0.$

  }


 \section{Conclusion}

The extraordinary Noether’s Theorem states that the existence of a group invariant transformation of the field
equations corresponds to a conservation flow. 
 Therefore such transformations can be
used  to recognize or select systems in gravity. The main purpose of this work was to search for quadratic forms of the conserved flows in the geometric background of  multi-scalar fields interacting  in their kinematic and potential parts in a spatially flat
FRW spacetime.  We have determined the interactions for which the  field dynamical system  is Liouville
integrable, according to the presence of dynamical symmetries.  Within the cosmological model, we assumed that the 2D metric $H_{AB}$ that characterizes the interaction in the field space
 is a space of constant non-vanishing curvature. Specifically, the metric space admits  three
Killing vectors which span the $SO(3)$ group. We used the  
 the existence of Noether symmetry contact transformations  to select the unknown form of the potential functions.  A detailed examination of the Killing tensors of the minisuperspace
 led us to the discovery of  two distinct cases  of the potential. The first admitted six quadratic conservation laws, and the second case produced a total of ten  quadratic conservation laws. These quadratic  integrals are in involution with the Hamiltonian function, i.e. by implication this property conveys that the field equations are integrable. Once we established the integrability of the system, the
cosmological solutions were found by quadrature for each case. Additionally, we first transformed the dynamical system into simpler but equivalent systems by constructing and applying a set of normal coordinates. In the first case, the scale factor of the universe appeared to be power-law, and for the case of the second potential, exponential-law.  Finally, some numerical simulations  were performed in order to study the Hubble, deceleration and equation of state parameter.

\textbf{Acknowledgments:} We acknowledge the
financial support from the National Research Foundation of South Africa (99279).


\end{document}